\documentclass{aa}  
\usepackage{graphicx}
\usepackage{psfig}
\usepackage{txfonts}
\def\fexvii{Fe\,{\sc xvii}}
\def\mrk{Mrk\,841}
\begin{document}
   \titlerunning{Warm absorber and warm emitter in Mrk 841}
   \authorrunning{A.L. Longinotti et al.}
   \title{High-resolution X-ray spectroscopy of the Seyfert 1 Mrk~841: insights into the warm absorber and warm emitter}

   \author{A.L. Longinotti \inst{1,2}, E. Costantini \inst{3},  P.O. Petrucci \inst{4}, 
          C. Boisson \inst{5}, M. Mouchet  \inst{5,6}, M. Santos-Lleo \inst{2}, G. Matt \inst{7},  G. Ponti \inst{6}, A. C. Gon\c{c}alves \inst{5,8}
          }

   \offprints{A.L. Longinotti}

   \institute{1 MIT Kavli Institute for Astrophysics and Space Research, 77 Massachusetts Avenue, NE80-6011, Cambridge, MA, 02139, U.S.A. \\
    2 ESAC - European Space Astronomy Centre,  P.O. Box 78, 28691 Villanueva de la Ca{\~n}ada, Madrid, Spain  \\
     3 SRON National Institute for Space Research, Sorbonnelaan 2, 3584 CA Utrecht, The Netherlands \\
 4 Laboratoire d'Astrophysique, UMR 5571, Universit\'e J. Fourier/CNRS, Observatoire  de Grenoble BP 53, F-38041 Grenoble cedex 9, France \\
 5 LUTH, UMR 8102, Observatoire de Paris, CNRS, Universit\'e Paris Diderot, 5 Place Jules Janssen, 92190 Meudon, France \\
 6  APC, UMR 7164, Universit{\'e}  Paris 7 Denis Diderot, 75205 Paris cedex 13, France \\
 7 Dipartimento di Fisica, Universit{\`a} degli Studi  Roma Tre, Via della Vasca Navale 84, 00146 Roma, Italy \\
8 CAAUL, Observat\'orio Astron\'omico de Lisboa, Tapada da Ajuda, 1349-018 Lisboa, Portugal
 }
 
%
          

   \date{Received ...;}

 
  \abstract
   {The Seyfert 1 galaxy Mrk~841 was observed five times between 2001 and 2005 by the  XMM-Newton X-ray observatory. The source is well known for showing  spectral complexity in the variable iron line and in the soft X-ray excess.
     }
   {The availability of multiple exposures obtained by  the reflection grating spectrometer (RGS) cameras  allows  thorough study of the complex absorption and emission spectral features in the soft X-ray band. This paper reports on the first study of Mrk~841 soft X-ray spectrum at high spectral resolution. }
   { The three combined exposures obtained in January 2001 and the two obtained in January and July 2005 were analysed with the {\small SPEX} software. }
   {We detect a two-phase warm absorber. A medium ionisation component  (log$\xi$$\sim$1.5-2.2~ergs s cm$^{-1}$) is responsible for a deep absorption feature in the unresolved transition array of  the Fe M-shell  and for several absorption lines in the OVI-VIII band, and a  higher ionisation phase with log$\xi$$\sim$3~ergs s cm$^{-1}$ is required to fit absorption in the NeIX-X band.  The ionisation state and the column density of the gas present moderate variation from 2001 to 2005 for both phases.  The high ionisation component of the warm absorber has no effect on the Fe K band. No significant velocity shift of the absorption lines is measured in the RGS data. Remarkably, the 2005 spectra show  emission features consistent with photoionisation  in a high-density (n$_e$$\ge$10$^{11}$~cm$^{-3}$) gas. A prominent  OVII line triplet was clearly observed  in January 2005  and  narrow radiative recombination continua (RRC) of OVII and CVI were observed in both 2005 data sets.  A broad Gaussian line  around 21.7~$\AA$ was also required to fit  all the data sets.  The derived radial distance for the emission lines seems to suggest that  the photoionisation  takes place within the optical broad line region of the source. 
    }
   {}


   \maketitle
%


\section{Introduction}
The model that  explains active galactic nuclei (AGN) as being  powered by accretion on a supermassive black hole surrounded by gas (Rees, 1984) is nowadays widely accepted. In the past decade, the advent of high-resolution X-ray spectroscopy has allowed  detailed study of ionised material flowing out along the line of sight within the central region (e.g. Kaspi et al. 2001; Kaastra et al. 2002). 
Multi-wavelength simultaneous observations have shown that  X-ray outflows and UV winds are tightly connected and that the absorption must take place in a material with multiple ionisation layers (Gabel et al. 2005). 
More than 50\% of the Seyfert~1 galaxies are characterised by this gas, which is mostly revealed in the X-ray band through detection of narrow absorption features from ionised elements, such as  carbon, oxygen, neon, iron (Creenshaw et al. 2003, Blustin et al. 2005).

The structure of the outflowing wind and its relation to the central engine where the AGN continuum radiation originates, are not entirely clear.
The absorbing gas seems to be  distributed diversely in individual sources,  its location spanning a scale of light days in, e.g., NGC~4051 (Krongold et al. 2007), to a few pc in, e.g., NGC~3783 (Behar et al. 2003).

Variability in the ionised absorbers can often offer a key to studying the change in the opacity of the gas in relation to the continuum emission.  For example,  long Chandra observational campaigns on bright Seyfert Galaxies  have yielded a variety of results on the response of the warm absorber
to the source luminosity fluctuations  (Netzer et al. 2002; Krongold et al. 2005).

More recent studies have shown that the presence of emission lines in the soft X-ray band brings additional information on the circumnuclear ionised gas. In some Seyfert 1s, emission lines  from highly ionised species can originate very close to or within the broad line region of the AGN  (Costantini et al. 2007, Longinotti et al. 2008), but only in sources with high signal-to-noise  has  it been possible to relate the X-ray emitter with the X-ray absorber (NGC~3783, Behar et al. 2003).

Mrk~841 is a bright Seyfert~1 galaxy at {\itshape z} = 0.0365  (V{\'e}ron-Cetty \& V{\'e}ron, 2001) that has been observed 
by several X-ray observatories in the past.
Spectral variability was reported in the {\it EXOSAT, Ginga} and {\it  ROSAT} data by  George et al. (1993), and by Nandra et al. (1995).
Evidence for a broad iron emission line was found in the {\it ASCA} spectra reported by Nandra et al. (1997).  Weaver et al. (2001) have reported marginal evidence for line variability. The analysis of broad band {\it BeppoSAX} data revealed a soft X-ray excess and a Compton reflection component (Bianchi et al. 2001, 2004), although these authors found that warm absorption provided an equally good description of the soft X-ray data. Evidence for a moderately deep OVII absorption edge was also reported in {\it ASCA} data by Reynolds (1997).
 {\it XMM-Newton} data at CCD resolution highlighted the presence of a variable and complex narrow Fe~K$\alpha$ line (Petrucci et al. 2002, Longinotti et al. 2004,  Petrucci et al. 2007).   Besides the study of Fe~K band, the later paper focused on the analysis of the soft excess in Mrk~841 by testing two competing  models. On one hand, the soft excess could be modelled by reflection off a photoionised accretion disc whose spectrum is relativistically blurred and smeared  (Crummy et al. 2006). A valid alternative model was proposed  by Gierlinski \& Done (2004) who explained the soft excess as arising from  ionised absorption in a relativistically smeared wind. According to Petrucci et al. (2007), both models were statistically indistinguishable in the EPIC data of Mrk~841.

 This paper presents the very first analysis of the high-resolution soft X-ray spectra of Mrk~841, and it makes use of all the available {\it XMM-Newton} data sets,
  i.e., the five observations previously studied by Petrucci et al. (2007). 
 
\begin{table}            
\caption{\label{tab:log} {\it XMM-Newton} observation log  for Mrk~841}                  
\begin{tabular}{c c c c c}
\\      
\hline\hline                
 Date          & OBSID &  Exp &  $^1$count rate   & $^2$F$_{0.3-2 keV}$  \\    
(yyyy/mm/dd)  &        -    & (ks)   & Cts/s      & (10$^{-12}$ cgs)    \\
\hline\hline 
\\ 
 2001/01/13   & 0112910201 &  10 & 0.63$\pm$0.01 &  26.6$\pm$1.0  \\
 2001/01/13  & 0070740101 & 12 &   0.72$\pm$0.01 & 26.6$\pm$1.0    \\
 2001/01/14  & 0070740301 & 14 &    0.71$\pm$0.01 & 26.6$\pm$1.0   \\
\\
 2005/01/16  & 0205340201 &  45 &   0.19$\pm$0.01 &   7.5$\pm$0.5  \\
\\
 2005/07/17  & 0205340401 &  30   &  0.23$\pm$0.01 & 9.1$\pm$0.1   \\              
\hline                                  
\end{tabular}

$^1$ In RGS 2\\
$^2$ Not corrected for absorption; the 2001 flux refers to the global fit of 3 data sets  

\end{table}


\section{The XMM-Newton observations}
Mrk~841 was observed  by {\it XMM-Newton} 3 times in 2001 and twice 
in 2005 (Table \ref{tab:log}).
In this paper, we present only the high-resolution spectra gathered by the RGS instrument (den Herder et al., 2001).
 For more details  on the EPIC data and on the broadband spectrum, the reader is referred to Petrucci et al. (2007).
All the raw data were processed with SAS version 7.0.0.
Spectral files for source and background were produced by running the standard procedure 
\texttt {rgsproc}.
Response matrices were created  by the task \texttt{rgsrmfgen}.
The observation performed in January 2005 presents an anomaly in the illumination of CCD~2 (for both 
 RGS1 and RGS2).  The background subtraction was checked and confirmed as unaffected by this event.  
 
 The source varied significantly in flux and spectral shape over the time covered by {\it XMM-Newton} observations (see Table~\ref{tab:log} and see Petrucci et al. 2007
 for an extensive discussion on the variability properties of the source). 
 The soft X-ray flux decreased by a factor of $\sim$3.5  from 2001 to 2005, thus  it is necessary to consider these two  ``epochs"  separately.
 
 The three short exposures of 2001 were checked carefully to decide whether the six RGS1 and RGS2  could be combined in two spectra (one for each instrument). We fitted the spectra with the same model to check the agreement in the spectral shape and in the flux level. No significant variation is found in the spectral shape in the RGS band, but since the count rates for the 2001 exposures differ by a few percent (see Table~\ref{tab:log}), instead of co-adding the data, we decided to fit simultaneously the six spectra of 2001 assuming the same fitting model and normalising the data to the RGS1 in observation 0112910201. 

This approach is maintained throughout the paper, and we refer to the 2001 simultaneously fitted spectra  simply as  ``Jan 2001". 
 Since the exposures of 2005 are separated by 6 months, we checked whether they could be combined into a single one.
After fitting the two 2005 spectra with a power law, the spectral slopes were found to be different by $\sim$20\%, therefore the 2005 spectra were not co-added in the analysis.

In summary, the three epoch spectra are kept disjointed during the entire spectral analysis and are referred to as Jan 2001, Jan 2005, and July 2005 throughout the paper.

\section{RGS spectral analysis}

The RGS spectral analysis was carried  out by using the latest version of the fitting package {\small SPEX} (Kaastra et al. 1996)\footnote{See also http://www.sron.nl/spex}. Galactic absorption of column density N$_H$=2.7$\times$ 10$^{20}$cm$^{-2}$ was included in all the following spectral fits (Dickey \& Lockman, 1990). The galactic absorption was modelled with the {\small SPEX HOT} component with a temperature fixed to 5$\times$10$^{-4}$~keV. For all spectral models, we adopted Solar elemental abundances  (Anders \& Grevesse, 1989).

The first order spectra were rebinned of a factor of 3 over 
the range 7 to 35~$\AA$ and the C-statistics was adopted (Cash, 1979).
Errors are quoted at the 90\% confidence level for one interesting parameter.
Throughout the text and tables, the quoted wavelengths have been corrected for the cosmological redshift.

\subsection{The warm absorber}
\label{sec:wa}

\begin{figure}
\includegraphics[width=0.7\columnwidth,angle=90]{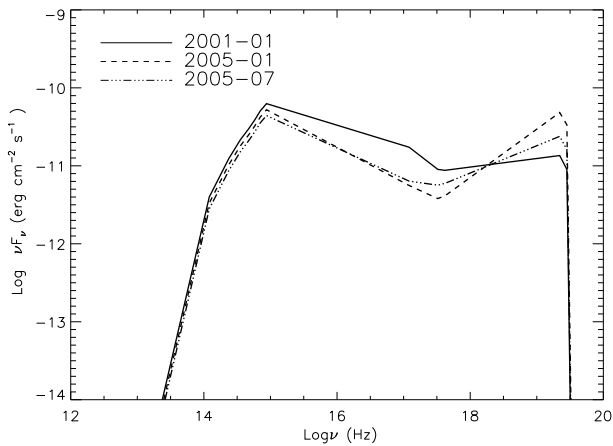}
\caption{Spectral energy distribution of Mrk~841 constructed from the {\it XMM-Newton} data in the optical-X-ray band and from the standard AGN radio-IR continuum included in {\small CLOUDY} (see text for more details).}
 \label{fig:sed}
   \end{figure}

\begin{figure}
\includegraphics[width=\columnwidth]{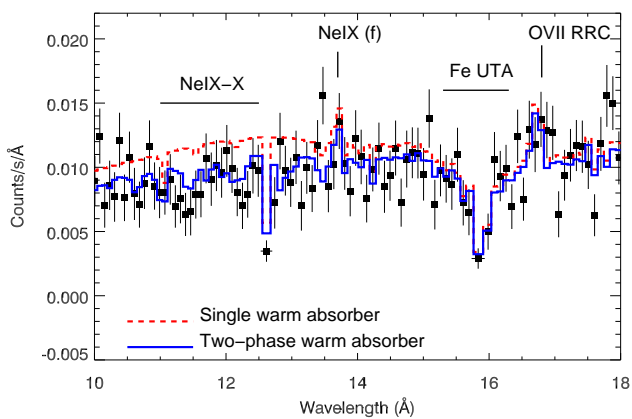}
\caption{ Jan 2005 RGS~2 data modelled with a single component (dashed red line)  and a two-component warm absorber (solid blue line).  The forbidden emission line of helium-like neon and the RRC of OVII,  which are included in both models (see Sect.~\ref{sec:emission_feat}), are labelled for clarity.}
 \label{fig:neon}
   \end{figure}

At first glance,  visual inspection of the fluxed spectra (i.e. the spectrum plotted in flux units rather than counts/sec) reveals the typical signatures of absorption from  
the so-called unresolved transition array of the Fe M shell in the 15-16~$\AA$ band (e.g. Sako et al. 2001). 
The feature appears to be prominent in Jan 2005, but it is also easily recognised in the other two data sets.
We checked  the residuals after fitting the spectra with a simple power law, confirming  absorption in all of them, although with different characteristics.
This may suggest variability in the ionised material along the three epochs and/or the presence of a multi-component warm absorber.
In order to compare the three spectra and test this hypothesis, we took the approach of initially fitting them with the same model.

We constructed a warm absorber model with the SPEX built-in model {\small XABS}. 
The ionisation balance was calculated by assuming  the spectral energy distribution (SED) of Mrk~841 as the ionising continuum spectrum. The SED was constructed with the simultaneous  optical-UV-to-X-ray fluxes extracted from the optical monitor and the EPIC instruments onboard {\it XMM-Newton}. 
 For the 2001 data set the photometric point in the U band (centred at 344~nm) was used. In Jan 2005 and July 2005, the source was observed with the optical grism, which produces a broader band spectrum. 
 The flux was extracted  at 344~nm for those data sets. At first approximation the complex X-ray unabsorbed broad band continuum can be modelled as a broken power law with an energy break at about 1.5~keV (as derived by a phenomenological fit to the EPIC data, see also Fig.~5 of Petrucci et al. 2007). For each of the three data sets, we took the soft power-law index from the RGS and the harder energy power law index from Petrucci et al. (2007). The very-low energy portion of the SED (radio to IR) is the same as the standard AGN continuum used in {\small CLOUDY} (Ferland et al. 1998). The three SEDs are plotted in Fig.~\ref{fig:sed}.
Then, we constructed a model for the continuum consisting of an underlying single power law absorbed by warm ionised gas and  applied it to the 0.35--1.77~keV data (i.e. the RGS band).

\begin{table*}   
\caption{\label{tab:wa}  Best-fitting parameters for the two-phase warm absorber.}      
\centering                  
\begin{tabular}{c c |c c c |c c c |c}
\hline\hline                
Obs &  $\Gamma$$_{soft}$ &  Log$\xi$   &  N$_H$  &  {\it v} &  Log$\xi$   &  N$_H$  & {\it v} &  Cstat/d.o.f. \\    
    -   &        -       & (ergs s cm$^{-1}$) & 10$^{21}$ cm$^{-2}$  &  km~s$^{-1}$ & (ergs s cm$^{-1}$) & 10$^{21}$ cm$^{-2}$ & km~s$^{-1}$ &     -        \\
\hline\hline 
\\ 
Jan 01   &  2.60$\pm$0.07  & 2.2$\pm$0.2   & 1.2$\pm$0.5   &  $<$1600  &  3.3$^{+ 0.8}_{-0.4}$  & $<$17.0  & 1000$\pm$460 & 3082/2725  \\

Jan 05   &    2.41$\pm$0.07   &  1.5$\pm$0.1    & 2.3$\pm$0.5  & $<$ 100   &  3.2$\pm$0.1  & 30.0$^{+110}_{-20.0}$  &  $<$100  & 1958/1435 \\

July 05   &   2.07$\pm$0.07   &  1.7$^{+0.3}_{-0.2}$   & 3.9$^{+2.6}_{-1.6}$  & $<$1000 &   2.8$\pm$0.3  & 7.6$^{+18}_{-5.5}$ & $<$1000  & 1855/1437   \\    
\\           
\hline                                  
\end{tabular}
\end{table*}   

The column density N$_H$ and the ionisation parameter $\xi$ of the gas are free parameters.
The ionisation parameter is defined as $\displaystyle{\xi=\frac{L}{n_eR^2}}$, where {\it L} is the ionising luminosity in 1-1000~Rydberg, {\it n$_e$}  the electron density of the gas, and {\it R}  the distance of the gas itself from the central source of radiation.
The {\small XABS} model accounts for the width of the absorption lines through  the  velocity dispersion parameter.  It has been shown in several Seyfert  galaxies that UV and X-ray  absorbers  originate in the same outflowing gas, thus UV observations can in general provide  a reliable limit on the X-ray line broadening (Arav et al. 2007, Costantini et al. 2007).  Since no constraint on the velocity dispersion of the UV lines is readily available in the literature on Mrk~841, this parameter was initially fixed to an average value of  20~km~s$^{-1}$. When it was left free in the fitting process, it tended to decrease to zero in all the spectra, thus it was eventually kept fixed to 20~km~s$^{-1}$.
Other free parameters in the model  are the photon index of the power law in the RGS band and the outflowing velocity of the warm absorber.
The fit statistic improves significantly with the addition of the {\small XABS} warm absorber model with respect to the power-law model, but the fit is still not completely satisfactory for all the three epochs. 

A close inspection of the spectra shows that absorption features are visible as negative residuals in the neon band (12--14~$\AA$). The single warm absorber component is not sufficiently ionised to account for absorption in this band.  Therefore,   a second warm absorber with a higher ionisation level  was tested by including in the model a second  {\small XABS} component. 
The addition of the second warm absorber leads to a fit statistic improvement by $\Delta$C=75 and $\Delta$C=25 for 3 d.o.f.  respectively measured in Jan and July~2005 spectra.  The resulting best-fitting parameters are reported in Table~\ref{tab:wa}. The soft photon index refers to the power law fitted in the RGS band, i.e. 7-35~$\AA$. The quoted parameters and C-statistic refer to the model including the emission features in Table~\ref{tab:emlines}. The improvement is less significant in Jan~2001, but this is not surprising since the column densities in Table~\ref{tab:wa} indicate that the absorber may be shallower at this epoch.
The effect of the second warm absorber  is shown in Fig.~\ref{fig:neon} for Jan~2005: the Fe UTA is clearly well fitted by the medium ionisation phase, whereas the absorption features of NeIX 1s-3p at 11.556~$\AA$, and NeX Ly$\alpha$ at 12.132~$\AA$  require a more highly  ionised plasma to be accounted for.

\begin{figure*}
\includegraphics[width=18cm]{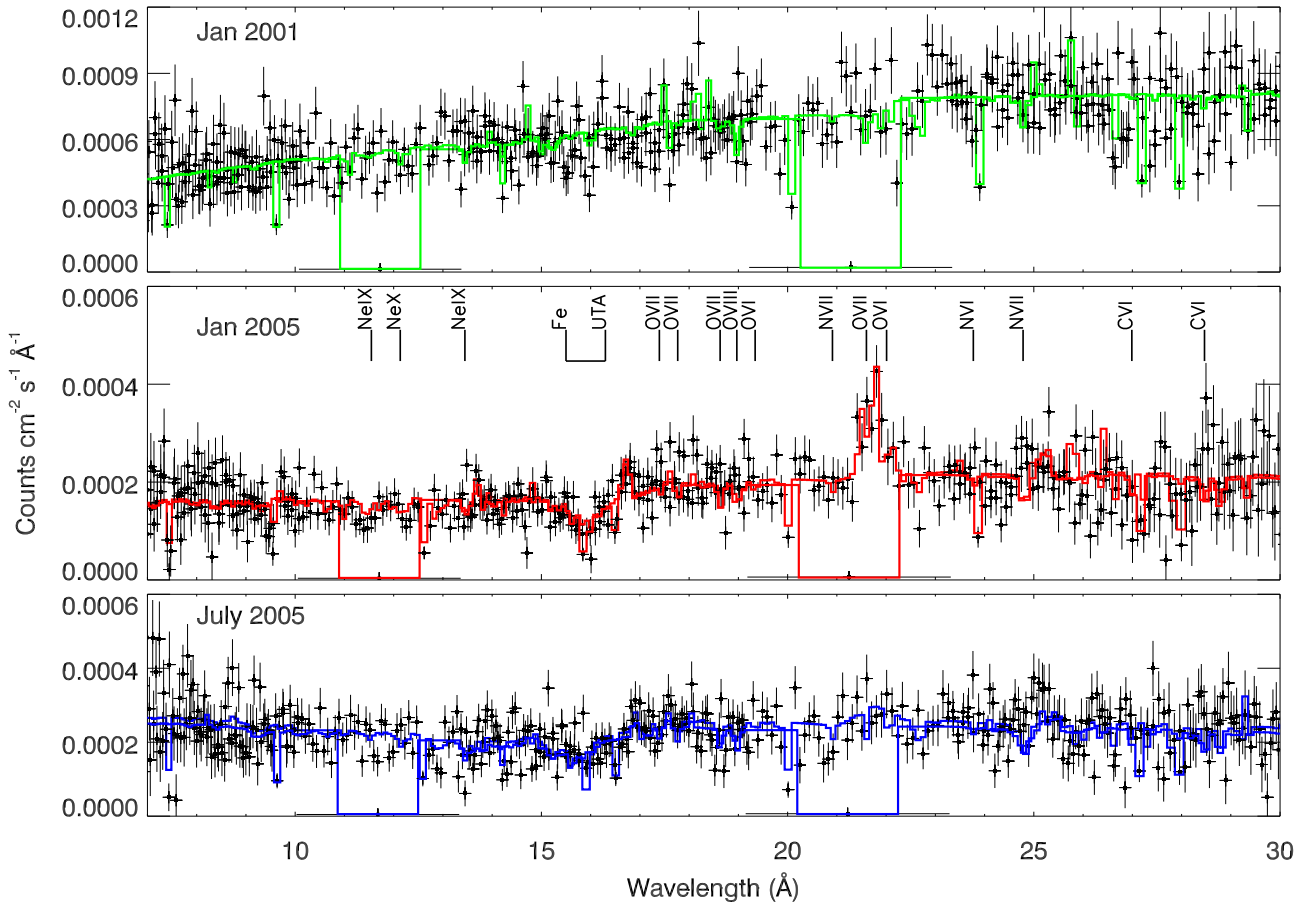}
\caption{From top to bottom: RGS rest-frame spectra of Jan~2001 (for plotting purpose only obsid 0070740101 is shown), Jan~2005 and July~2005 fitted with the best-fitting model including the components in Tables~\ref{tab:wa} and \ref{tab:emlines}. For plotting purposes, the spectra are re-binned by a factor of 9 and plotted in the 7-30~$\AA$ band, where most of the features of interest fall. The labels mark the ionic species with the highest contribution in our warm absorber model.}
 \label{fig:bestfit}
   \end{figure*}

 The two-phase warm absorber provides a good fit to the data as can be seen in  Fig.~\ref{fig:bestfit}.  
The main absorption features are identified with transitions of OVI, OVII, OVIII  in the 18-23~$\AA$ band,  Fe UTA  at 15-16~$\AA$ and Neon IX-X in the 12--14~$\AA$.  
No significant velocity shift is detected in any of the two warm absorbers (see Table~\ref{tab:wa}), except for the high ionization phase in Jan~2001.  The upper limits are consistent with outflowing absorbers blueshifted by several~$\times$10$^2$ km~s$^{-1}$, as commonly observed in Seyfert~1 Galaxies (e.g. McKernan et al. 2007).

Even with  the addition of the second warm absorber, the fit of the Jan~2005 spectrum is not completely satisfactory, since positive residuals  are still present in the OVII band (21-22~$\AA$, see top panel of Fig.~\ref{fig:triplet}).
The analysis of these residuals in emission is reported in the following section.

\subsection{Emission features in the RGS spectra}
\label{sec:emission_feat}
\subsubsection{Narrow emission lines}
\label{subsec:lines}

\begin{figure}[t]
\begin{center}
\includegraphics[height=.9\columnwidth,width=0.65\columnwidth,angle=-90]{12925_fg4a.ps}
\includegraphics[width=\columnwidth]{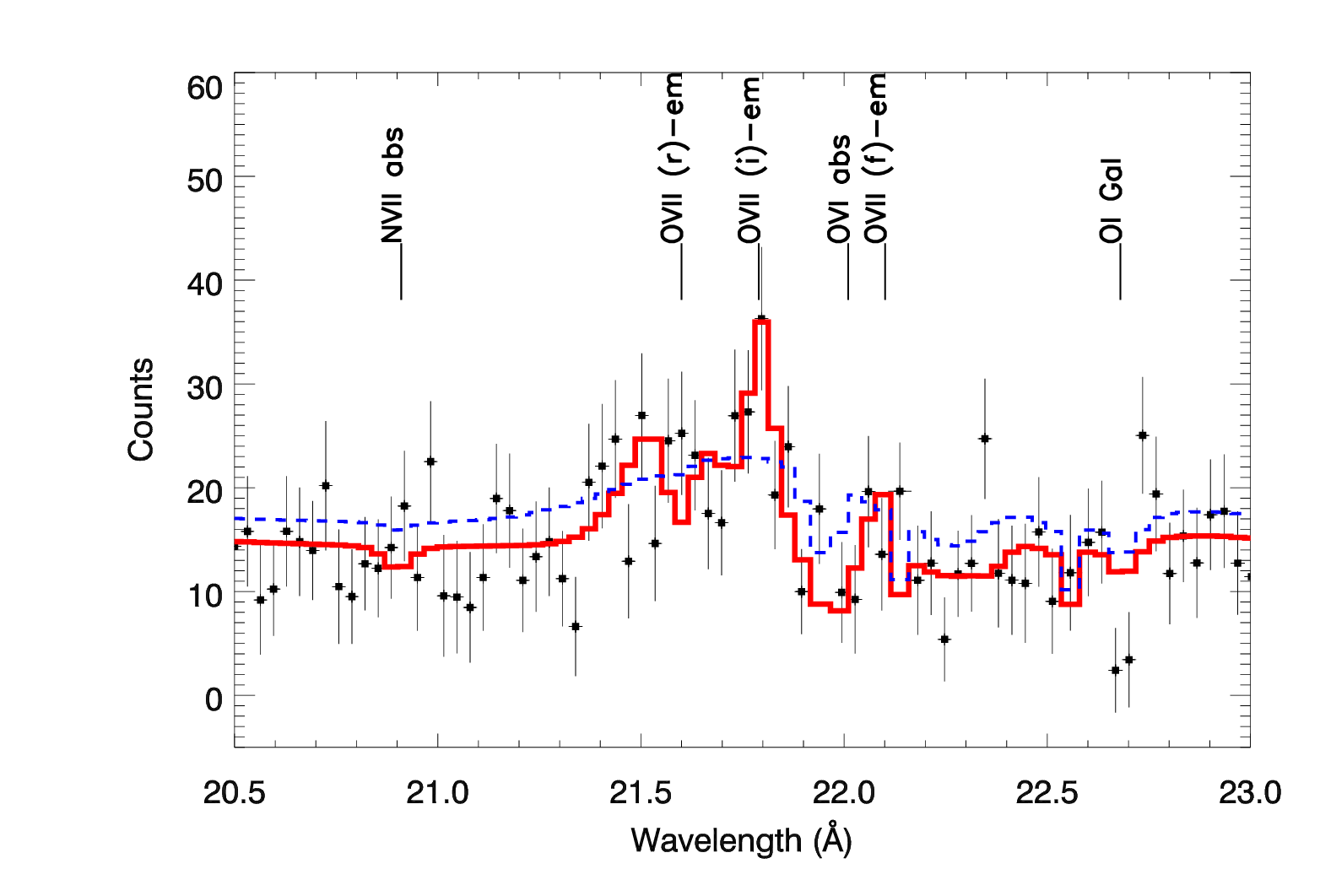}
\end{center}
\caption{Top: Residuals of the continuum model in the OVII band.  Bottom: Zoom on the OVII triplet in Jan 2005. Line parameters are reported in Table 3. The  solid red line corresponds to the best-fitting model (including the emission lines in Table~\ref{tab:emlines} and the underlying warm absorber in Table~\ref{tab:wa}). The dashed blue line shows the underlying OVII broad Gaussian line after removing  the contribution of the narrow lines triplet and the warm absorber.}
\label{fig:triplet}
\end{figure}

We start by fitting  Jan~2005 with three narrow-line components  at the wavelengths corresponding to the rest frame positions of the  resonance, intercombination, and forbidden lines ({\it r, i, f}) in the  OVII line triplet at 21.600,  21.790, and 22.101~$\AA$, respectively. The analysis of emission lines in this portion of the spectrum is complex because of the superposition of the absorption lines imprinted by the warm absorber. The peak wavelengths of the three lines are kept frozen to the laboratory values. Since the {\it r} transition has a large oscillator strength in the absorption component, most of the emission line at this wavelength will be re-absorbed, making a precise measure of its flux somewhat difficult.   
 The {\it i} and {\it r } components are detected, while only an upper limit is found for the {\it f} line.  Less prominent positive residuals are present in the NeIX band, hence we searched for emission lines also in this portion of the spectrum.  Only the NeIX forbidden component at 13.699~$\AA$ is detected  (the neon band is shown in Fig.~\ref{fig:neon}).
The emission lines fluxes and significances in terms of C statistic improvement for one free parameter, are reported in Table~\ref{tab:emlines}. The structure of the line triplet fitted in Jan~2005 data is plotted in Figure~\ref{fig:triplet} (thick red line).

We model the spectra of Jan~2001 and July~2005 with the same emission line components included for Jan~2005.  Not surprisingly, given that the continuum flux is higher in these two data sets (see Table~\ref{tab:log}), we could measure only the upper limits on the line fluxes, which are generally consistent with Jan~2005 spectra, except for the OVII resonance line  in July~2005 that may be severely underestimated.

\subsubsection{Radiative recombination continua} 
The spectrum of Jan 2005 displays an additional emission feature around 17~$\AA$, consistent with the position of the radiative recombination continuum (RRC) of OVII (see Fig.~\ref{fig:neon} for a close-up of this feature).  A closer look at longer wavelengths suggests that other RRC components from Carbon may give significant contribution in the positive residuals of the data.   
In our analysis, these features are fitted with the {\small RRC} model in {\small SPEX}. The width of the RRC profiles is a very sensitive indicator of the electron temperature in the recombining  plasma (Liedahl \& Paerels 1996, Liedahl 1999). Our data suggest the presence of a narrow RRC at the OVII wavelength (see Fig.~\ref{fig:neon}), indicative of temperature of few$\times$10$^{4}$~K. 
Therefore, we fit the RRC emission measure\footnote{The emission measure corresponds to n$_e$n$_{ion}$V, where n$_e$ is the gas electron density, n$_{ion}$  the density of the parent ion that recombines to the ground state, and  V the emitting volume}  of OVII and CVI as free parameters assuming    
  an electron temperature corresponding to {\it kT}= 3~eV, i.e. T=3.5$\times$10$^{4}$~K. If the temperature is left free to vary in the fit, we find {\it kT}=2.8$^{+2.9}_{-1.4}$~eV.

  The improvement in the fit statistic for adding the RRC component in Jan 2005 is $\Delta$C=23 for 2 degrees of freedom. 
  The OVII~RRC  seems to be slightly redshifted from its laboratory position; we  applied a small redshift of {\it z}=0.001 to match the observed excess in the data. The same was done for the C~VI RRC around 25~$\AA$, where the shift is less evident due to the lower signal-to-noise of our spectra in this band.
 As described in the previous section for the narrow emission lines, we included  the RRC in the model for all the three spectra and  reported the best-fitting parameters in Table~\ref{tab:emlines}.

\subsubsection{Broad oxygen line}

Once the narrow lines are included in the fit, the residuals around the OVII triplet in Jan~2005 still show some structure in excess, particularly  on the left side of the resonant line. To account for these residuals, a broad Gaussian line was added in the model with width and wavelength free to vary. 
The resulting best-fitting parameters for this component are $\lambda$=21.75$^{+0.23}_{-0.36}$~$\AA$, FWHM=0.60$^{+0.52}_{-0.35}$~$\AA$, and flux=0.64$\pm$0.53$\times$10$^{-4}$~photons~cm$^{-2}$~s$^{-1}$. The addition of the broad line component leads to an improvement of $\Delta$C stat=18
for 3 degrees of freedom, corresponding to a significance level higher than 99\%.  Fig.~\ref{fig:triplet} shows the separate contribution of the Gaussian profile to the best fit model for the OVII line triplet.        
 For simplicity,  the broad line at 21.7~$\AA$ will only be referred to as ``OVII line", without identifying a specific transition.
 The same line component was included in July 2005 yielding the following 
best-fitting parameters: $\lambda$=21.71$\pm$0.23~$\AA$,  FWHM=0.54$^{+0.57}_{-0.44}$~$\AA$ and flux=0.51$\pm$0.38$\times$10$^{-4}$~photons~cm$^{-2}$~s$^{-1}$. Although the significance of this component in July 2005 (around 90\%) is lower than in Jan 2005, the generally good agreement in the best-fitting  parameters for the two spectra seem to support the presence of the same line in July 2005.
In Jan 2001 data, we added the broad  component as we did for the narrow lines. Since the higher continuum flux prevents any determination of the line properties, to measure the line flux,  the position and width of the Gaussian component have been fixed to the average value found for the 2005 epochs (see Table~\ref{tab:emlines}).
An upper limit of $<$0.85~photons~m$^{-2}$~s$^{-1}$  (consistent with the 2005 measurements) was found for the broad component in 2001.

\begin{table*} 
\centering  
\caption{\label{tab:emlines}  Emission lines parameters and emission measure for RRC in the three epochs.}      
\begin{tabular}{c | c c c c c |c}
\hline\hline   & & & & & &      \\      
Obs  &  Line  &  $\lambda$  &  Flux  &  $\Delta$C$^1$ &  FWHM &  RRC EM  \\    
 -   &   -    &   ($\AA$)   & (10$^{-4}$ph~cm$^{-2}$~s$^{-1}$)   & -  & ($\AA$)  &   (10$^{58}$~cm$^{-3}$) \\
\hline\hline   & & &  & & & \\
               & OVII  (r)  &  21.600    &  $<$   0.42  &   - & 0-width   &              \\
               & OVII  (i)  &  21.790    &  $<$   0.15  &   - & 0-width  & OVII $<$450 \\
 Jan 01        & OVII  (f)  &  22.101    &  $<$   0.43  &   - & 0-width   & CVI$<$4300  \\ 
               & NeIX  (f)  & 13.699     &  $<$   0.12  &   - & 0-width    &             \\ 
& & & & & & \\
               & OVII (gau) &  21.7    &  $<$   0.85  &   - & 0.55 &            \\
    & & & & & &    \\                    
\hline       & & & & & &      \\

            &  OVII  (r)  &  21.600     & 0.24$\pm$0.16             & 17  & - &     \\
            &  OVII  (i)  &  21.790     & 0.26$^{+0.26}_{-0.16}$    & 16  & - &  OVII=1500$\pm$800 \\
Jan 05      &  OVII  (f)  &  22.101     & $<$0.20                   &  -  & - &  CVI=2500$\pm$1800  \\ 
            &  NeIX (f)   & 13.699      & 0.10$\pm$0.05             & 10  & - &   $\Delta$C stat=23   \\ 
 & & & & & &    \\
& OVII (gau)& 21.75$^{+0.23}_{-0.36}$ & 0.64$\pm$0.53             & 18  & 0.60$^{+0.52}_{-0.35}$ &  \\
    & & & & & &  \\   
\hline        & & & & & &   \\

           & OVII (r)    &  21.600   &  $<$0.04    &  -  & - &     \\  
           & OVII (i)    &  21.790   &  $<$0.25    &  -  & - & OVII=1860$\pm$1000 \\  
July 05    & OVII (f)   &   22.101   &  $<$0.23    &  -  & - & CVI=2500$\pm$2300 \\  
           & NeIX (f)   &   13.699   &  $<$0.15    &  -  & - &  $\Delta$C stat=15  \\           
 & & & & & &    \\
           & OVII (gau) & 21.71$\pm$0.23 & 0.51$\pm$0.38 &  6 &  0.54$^{+0.57}_{-0.44}$   &       \\

    & & & & & &     \\   
      \hline\hline                           
\end{tabular}

$^1$When $\Delta$C stat $\le$3, the upper limits on the line flux are reported \\
\end{table*}



\section{Discussion}
The analysis of the RGS data of Mrk~841 presented herein provides a considerable amount of information.
A large contribution from a warm absorber consisting of two ionisation components has been revealed and most interestingly, it proves to be relatively stable on a time scale of 4 years. 
In addition,  emission features from highly ionised species of neon, oxygen, and carbon have been ascertained.
 These topics will be addressed separately in the following sections. The discussion is based on the results from the global best fit (i.e. absorption+emission), which is reported in Tables~\ref{tab:wa} and \ref{tab:emlines}.


\subsection{A two-phase warm absorber in Mrk 841}
The properties of the ionised absorber emerging from the spectral fits described in Sect.~\ref{sec:wa}, indicate that there are two distinct levels of ionisation in the absorbing gas. 
A medium-phase warm absorber characterised by log$\xi$ in the range 1.5--2.2~ergs~s~cm$^{-1}$ and column density of the order of 10$^{21}$~cm$^{-2}$ is responsible for transition in the Fe UTA and in the oxygen band. 

The availability of spectra from different epochs may allow a variability study 
of the absorber physical conditions, particularly considering that because the Fe UTA is the result of a blend of absorption lines arising from several iron ions, its shape and position are very sensitive to changes in the ionisation balance of the gas and, ultimately, to luminosity variations of the source (Behar et al., 2001).
The errors in Table~\ref{tab:wa} show that $\xi$ and N$_H$ are generally  consistent through the three data sets, although 
there seems to be a decrease in ionisation accompanied by a moderate increase of the column density going from 2001 to Jan~2005. 
   Indeed, the Fe UTA is much more evident in both 2005 spectra because the lower ionisation state observed at this epoch is able to produce a high number of transitions in Fe~XV-XVII in the 15-16~$\AA$ region.  
According to our model, absorption from  OVI and OVII is also strong in this source.
  These absorption lines are not easily recognisable, especially  in Jan~2005 spectrum,  due to the presence of the OVII emission line triplet. The OVI-OVII absorption is likely associated to the gas  component in a medium ionisation state.

The higher ionisation phase (log$\xi$$\sim$3~ergs~s~cm$^{-1}$) is instead responsible for imprinting absorption at shorter wavelengths mainly in the neon band around 12--14~$\AA$. This component is characterised by a column density of the order of 10$^{22}$~cm$^{-2}$, much higher than the one measured in the medium phase. In principle, warm gas at this ionisation level can produce absorption lines in the Fe~K band around 6--7~keV, which may play a key role in shaping the Fe~K alpha line profile (e.g.  Revees et al. 2004, Turner \& Miller, 2009).  Indeed, Longinotti et al. (2004) and Petrucci et al. (2007)  reported  peculiar behaviour in the Fe~K$\alpha$ line in their previous analysis of Mrk~841, but none of these works included  detailed modelling of the warm absorber. 
 The detection of a high ionisation warm absorber in the RGS data prompted us to test for absorption effects on the Fe~K line.
  This check was performed by simply applying our best-fitting model to the EPIC-pn data of Jan~2005, i.e., the spectrum with the highest column density in the high ionisation phase (Table~\ref{tab:wa}). 
 With the parameters frozen to the values of  Table~\ref{tab:wa},  we found that this warm absorber component is not deep enough for introducing any additional curvature around the Fe line, or for modifying the line profile itself.

\subsection{Density of the warm absorber}
\label{subsec:rec_density}
In about four-years time, the low-energy flux of \mrk\ decreased by a factor of  3.5
(Table~1). In Jan~2005, when the source was fainter, we measured also 
a lower ionisation parameter for the cold phase, 
while the column density stayed constant within the errors (Table 2). Given the
 uncertainty on the outflow velocity, we assume here that we detected the same gas
material in both 2001 and 2005, albeit with different ionisations.

We estimated a lower limit for the density of such gas, assuming a 
recombination time of 1422 days (4 years time  between Jan~2001 and Jan~2005),  i.e. t$_s$=$\displaystyle\frac{t_o}{1+z}$ in the source rest frame. We considered all the Fe ions detected in the spectrum.  Each ion has its own
recombination time, which is inversely related to the gas density n$_e$, as reported in Bottorff et al. (2000) and Detmers et al. (2008):

$\displaystyle t(X_i)=\left(\alpha_r (X_i)n_e \left[\frac{f(X_{i+1})}{f(X_i)} - \frac{\alpha_r(X_{i-1})}{\alpha_r(X_i)} \right]\right)^{-1}$   
   
 where $t(X_i$) is the recombination timescale of the ion $X_i$, $\alpha$(X$_i$)  is the recombination rate from ion $X_{i=1}$ to ion $X_i$, and 
$f(X_i$) represents the fraction of the element {\it X} at the ionisation state {\it i}.

This relation  allows us to put further constraints on the lower limit of the density. 
We used the values  predicted by the best-fit global modelling of the
absorber, as the quality of the data does not allow a line-by-line
fitting on the Fe ions to be performed, especially in the crowded region of the iron UTA. The
associated error on the ionic column densities is therefore the scaled value of the
error on the total  $N_{\rm H}$. 

The gas density is a function of the recombination rates, which are in turn a
function of the gas temperature $T_e$, before recombination (Bottorff et al. 2000, Detmers et al. 2008). 
The value of $T_e$ is linked to the
ionisation parameter $\xi$ via the ionisation balance. The error on $\xi$
(Table~2),
rather than the error on individual ionic column densities, has the major 
weight in the density determination. The radiative recombination rates 
for iron were analytically determined using the expression and
recombination coefficient provided in Woods et al. (1981).

\begin{figure}
\begin{center}
\includegraphics[angle=90,width=9cm,height=6.5cm]{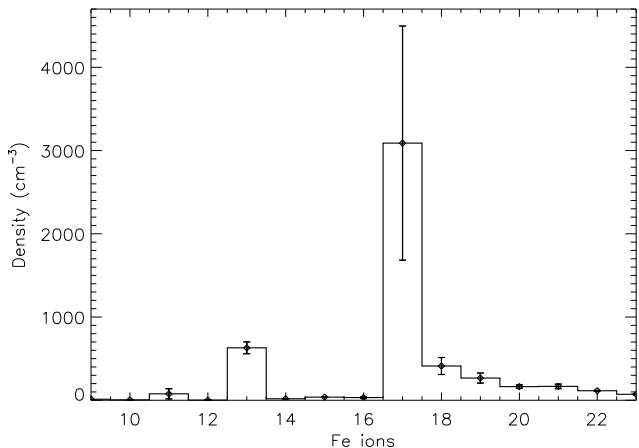}
\end{center}
\caption{\label{f:fe_ions} Electron density in the warm absorber estimated using the recombination of the iron ions produced
by the gas with log$\xi=2.2\pm0.23$ and $N_{\rm H}\sim1.2\pm0.5\times10^{21}$\,cm$^{-2}$.}
\end{figure}

In Fig.~\ref{f:fe_ions} we show the density lower limits for the Fe ions. The
ion that provides the higher value for $n=3.1\pm1.4\times10^{3}$\,cm$^{-3}$ 
is \fexvii.

\subsection{On the origin of the narrow emission lines and radiative recombination continua}
 The following discussion is based chiefly on the findings for Jan~2005 spectrum, where the emission line components are detected with highest significance.
  The narrowness of the RRC profiles is consistent with an electron temperature of a few~$\times$10$^{4}$~K, and it seems to indicate that photoionisation from the central source is the primary mechanism acting on the surrounding gas (e.g. Kinkhabwala et al. 2002, Armentrout et al. 2007).
 
The relative intensity of line triplets in He-like ions provides an excellent diagnostic of the physical properties of the emitting plasma 
(Porquet \& Dubau, 2000 and reference therein). In general, it has been observed that photoionisation processes in Seyfert Galaxies give rise to line triplets dominated by the forbidden transition (e.g. NGC~1068, Kinkhabwala et al. 2002). 
 If we consider the line fluxes  of the OVII triplet in Mrk~841,  we note that their ratio  is not typical of photoionisation,  the less prominent line being the forbidden transition,  as is also evident  from Fig.~\ref{fig:triplet}. 
The intercombination component at 21.790~$\AA$ is as strong  as the resonance one, if not higher:  the presence of a bad pixel  between the {\it r} and {\it i} lines does not allow  precise reconstruction of  the {\it r} emission line.
 On the contrary, we measure an upper limit for the forbidden component. Since this line is not affected by instrumental effects and is generally expected to be dominant in AGN spectra,  this measurement provides a stringent limit to be considered in the discussion of the line ratios.

 The {\it R} ratio $\displaystyle\frac{\it f}{\it i}$ provides a compelling probe of the gas density in photoionised plasma (see Porquet \& Dubau, 2000). 
  The low value of this ratio (R~$<$0.76) in Mrk~841 points to emission from a  gas at very high density, i.e.  {\it n$_e$} $\ge$~10$^{11}$~cm$^{-3}$. This result partly recalls the case of the Seyfert~1 Galaxy Mrk~335 that displayed a photoionised OVII triplet with a dominant  {\it i} line during a particularly low flux state (Longinotti et al. 2008). 

The {\it G} ratio $\displaystyle\frac{\it f+i}{\it r}$ instead is often used as an estimator of the electron temperature in the gas.
 We can estimate an upper limit on the {\it G} ratio by taking  the upper limits on the three components of the triplet  into account (i.e., the upper error bars for the two lines for which they are available and the upper limit on the {\it f} line).
In this way, the {\it G} ratio  is estimated to be less than 1.8.
This limit is somewhat intermediate between a pure collisional and a pure  photoionisation-dominated plasma (see Fig.~7 in Porquet \& Dubau, 2000), possibly indicating that mixed processes are at work.
We note some discrepancy between  the temperature of the recombining plasma inferred in this work by the RRC width ($\sim$10$^{4}$~K) and  the temperature prescribed for the hybrid plasma by the calculations reported in Porquet \& Dubau (2000). However, as highlighted  for Mrk~3 (Sako et al. 2000)  and  NGC~4151 (Armentrout et al. 2007), the intensity of resonance lines in He-like triplets could be enhanced by photo-excitation mechanisms induced by the AGN, which leads to a decrease in the  observed {\it G} ratio (Godet et al. 2004).
In principle, this effect could be tested in our data  by accounting for microturbulence in the resonance line (see Armentrout et al. 2007), but we recall that the velocity dispersion was undetermined even for the warm absorber lines (less than 20~km~s$^{-1}$, see Sect.~\ref{sec:wa}), thus  not easily  determined  with the present instrumental resolution. 

\subsection{One photoionisation region for the warm absorber and the warm mirror?}

While certainly very interesting, the soft X-ray properties in Mrk~841 are not unique because several AGNs show features from ionised plasma both in absorption and emission e.g. NGC~4151, (Armentrout et al. 2007, Kraemer et al. 2005) and NGC~3783 (Behar et al. 2003). For some cases, attempts are being made to ascribe the two processes to the same photoionised plasma (e.g. NGC~4051, Nucita et al. submitted). The main unknown is the geometry of the circumnuclear gas, which should  act as a ``warm mirror" for scattering the emission lines and, at the same time, it should intercept the observer's line of sight for producing warm absorber features.

 For a large number of X-ray obscured Seyfert galaxies, there are several indications that the X-ray photoionisation region where the bulk of soft X-ray line emission takes place, is spatially coincident with the optical narrow line region (NLR) (see Bianchi et al. 2006 and Guainazzi \& Bianchi 2007). Seyfert~1 objects, with unobscured X-ray spectra, offer instead a direct view of the continuum source and the surrounding high density gas giving rise to optical broad lines (i.e. the broad line region (BLR)),  therefore introducing more diversity in determining the distribution of the hot plasma.
 As an example, in the Seyfert 1 NGC~4051 (Pounds et al. 2004) and NGC~5548 (Kaastra et al. 2002, Detmers et al. 2008), the gas density inferred from the observed emission lines  indicates an origin in the NLR, whereas  in  Mrk~335, the high density line-emitting gas  (n$_e$=10$^{9-11}$~cm$^{-3}$) was located within the source BLR (Longinotti et al. 2008).

The present study of Mrk~841 with the detection of an important warm absorber component and  photoionised emission lines, provides the opportunity to compare and, possibly, link their properties.
 The  ionisation parameter of the photoionised gas $\xi$=$\displaystyle\frac{L}{n_eR^2}$  depends on the plasma electron density, its  radial distance from the source of radiation, and the ionising luminosity. An average  luminosity of 10$^{45}$~ergs~s$^{-1}$  was estimated from the SED in Fig.~\ref{fig:sed} by integrating the data  of Jan~2001 and Jan~2005 over 1-1000~Rydberg.

In Sect.~\ref{subsec:rec_density}, we derived a lower limit for the density of the warm absorber ($>$10$^3$~cm$^{-3}$), which can loosely constrain the distance to $<$ few tens of pc. 
Nevertheless, this constraint can be refined by using the limit on the density inferred by the OVII triplet (n$_e$$\ge$~10$^{11}$~cm$^{-3}$), under the assumption that the emission lines originate within a gas at the same ionisation state of the warm absorber gas.  
Our two-phase model for Jan~2005 yields log$\xi$$\sim$1.5 and 3.2~ergs~s~cm$^{-1}$ (see Table~\ref{tab:wa}). However, in the model of the gas with log$\xi$$\sim$3.2~ergs~s~cm$^{-1}$   the column density of OIX is much higher than the one of  OVII and OVIII, therefore such a highly ionised gas is unlikely to produce OVII emission.
Assuming the ionisation parameter of the low phase, the radial distance {\it R} of the ``warm emitter gas"  is then constrained to less than 1.6$\times$10$^{16}$~cm.

 The person who wishes to put this result  in the context of the source may consider that the optical  BLR lies at a radius of about 4.4$\times$10$^{16}$~cm, as roughly estimated from the FWHM of 5500~km~s$^{-1}$ in the optical lines (Boroson \& Green, 1992).  Thus, the ionised gas seems to encompass the BLR, analogously to Mrk~335 (Longinotti et al. 2008).
 We checked the agreement of the optical FWHM and the width of the OVII line triplet by fitting the three components with a Gaussian profile and freezing the rest of the model. We found the following FWHM for {\it r, i, f}, respectively: 400$\pm$60, $<$1400 and $<$5300~km~s$^{-1}$. While the velocities in the  {\it i} and {\it f} components  can still be reconciled with BLR, it is 
 hard to explain the apparent inconsistency of the resonant line with the rest of the triplet. 
We note, however, that the reconstruction of the {\it r} line is rather complex, as highlighted in Sect.~\ref{subsec:lines}.
A more sophisticated modelling of the emission lines in high signal-to-noise data would be needed to supplement our hypothesis.
 
A few words must be said on the line fluxes in the other two data sets for which only upper limits are available. 
 The apparent lack of strong emission lines  in Jan~2001 and Jan~2005 can be phenomenologically explained  as an effect of the increased continuum flux. 
 As a consequence,  the emission component is swamped and covered in the observed spectra. While it is hard to exclude intrinsic variability 
 in the emission line fluxes, especially since the continuum flux varies, there is no  compelling  evidence for variation in the present data.


\subsection{The broad OVII line in the RGS data}
The broad emission line around $\sim$21.7~$\AA$ is  detected with high significance in both  2005 observations. 
If fitted with a Gaussian profile, the FWHM corresponds to $\sim$8200$^{+7200}_{-4700}$~km~s$^{-1}$, for Jan~2005.
 Considering the large error bars, this velocity is consistent with the optical H$\beta$ FWHM of 5500~km~s$^{-1}$ reported by Boroson \& Green (1992). 
  Broad emission features in the soft X-ray spectra of Seyfert~1 Galaxies has been reported
 in the {\it Chandra}-LETG of Mrk~279 (Costantini et al. 2007),  NGC~5548 (Steenbrugge et al. 2005), NGC~4051 (Steenbrugge et al. 2009),  and in the {\it XMM-Newton}-RGS spectra of NGC~4051 (Ogle et al. 2004, Ponti et al. 2006).
 In  most of these cases, the line profiles were consistent with Gaussians, and in the first two sources,  the velocity width of the emission lines 
 was in agreement with measurements of UV lines in the BLR.

In the past,  it was proposed to model RGS spectral lines from H-like ions in the Seyfert~1 Galaxies
  of MCG-6-30-15 and Mrk~766  with a relativistic profile  (Branduardi-Raymont et al. 2001). 
We tried to convolve the OVII Gaussian emission line in Mrk~841 with a relativistic profile using the {\small LAOR} kernel (Laor, 1991) in Jan~2005  spectrum. 
 No good fit was found,  perhaps not surprisingly since the  data quality of Mrk~841 is not comparable to the 
long-integration RGS spectra of the two bright sources mentioned above, keeping the test for relativistic broadening from being viable in this source.

Considering the agreement of the OVII and H$\beta$ lines FWHM in Mrk~841, a BLR origin for the OVII broad line is favoured.  
 Unfortunately,  the lack of simultaneous UV spectral data for this source prevents us from performing a more detailed analysis of this feature and its relation to the multi-wavelength properties. 
  

  \section{Summary of conclusions}
In the following we summarise the main results of the first analysis of Mrk~841 high-resolution X-ray spectra: 
\begin{itemize}

\item Warm absorber: a two-phase warm absorber with log$\xi$$\sim$1.5-2.2~ergs~s~cm$^{-1}$ and log$\xi$$\sim$3~ergs~s~cm$^{-1}$  was observed in the three spectra from 2001 through 2005,  a moderate decrease in the ionisation state was observed going from the brightest flux state (2001) to the dimmer state (2005),  no outflow was detected in the data and the high ionisation component has no effect on the Fe~K band.

\item Emission features:  narrow OVII and NeIX  emission lines were detected in the Jan~2005 spectrum, the line ratio in the OVII triplet points to an origin in a gas at high electron density (n$_e$$\ge$10$^{11}$~cm$^{-3}$), the presence of narrow RRC from OVII and CVI is consistent with photoionisation processes, if it is assumed that the emission component and the warm absorber originate within a gas at the same ionisation state, the inferred distance for the emission lines is constrained within 10$^{16}$~cm from the central source.

\item Broad Gaussian: a broad Gaussian line centred at 21.7~$\AA$ was interpreted as emission from OVII,  the FWHM is fully compatible with that of the optical broad emission lines in Mrk~841.
\end{itemize}

\begin{acknowledgements}
      This paper is based on observations obtained with the {\it XMM-Newton} satellite,
an ESA science mission with instruments and contributions directly funded by ESA Member States and NASA.  We acknowledge support from the Faculty of the European Space Astronomy Centre (ESAC) and from the French GDR PCHE
for financially supporting the collaboration meetings needed to finalise this work.
GP thanks ANR for support (ANR-06-JCJC-0047).
 \end{acknowledgements}

\end{document}